\documentstyle[eqsecnum,aps,prd]{revtex}
\begin{document}
\thispagestyle{empty}
\thispagestyle{empty}
{\baselineskip0pt
\leftline{\large\baselineskip16pt\sl\vbox to0pt{\hbox{DAMTP} 
               \hbox{\bf\it University of Cambridge}\vss}}
\rightline{\large\baselineskip16pt\rm\vbox to20pt{
               \hbox{DAMTP-1999-85}
               \hbox{UTAP-340}
               \hbox{RESCEU-27/99}           
               \hbox{\today}
\vss}}%
}
\vskip15mm
\begin{center}
{\large\bf Stringy Probe Particle and Force Balance}
\end{center}

\begin{center}
{\large Tetsuya Shiromizu} \\
\vskip 3mm
\sl{DAMTP, University of Cambridge \\ 
Silver Street, Cambridge CB3 9EW, United Kingdom \\
\vskip 5mm
Department of Physics, The University of Tokyo, Tokyo 113-0033, 
Japan \\
and \\
Research Centre for the Early Universe(RESCEU), \\ 
The University of Tokyo, Tokyo 113-0033, Japan
}
\end{center}

\begin{abstract} 
We directly derive the classical 
equation of motion, which governs the centre of mass of 
a test string, from the string action. In a certain case, the equation is 
basically same as one derived by Papapetrou, Dixon and Wald for a test 
extended body. We also discuss the force balance using a stringy probe
particle for an exact spinning multi-soliton solution of 
Einstein-Maxwell-Dilaton-Axion theory. 
It is well known that the force balance condition 
yields the saturation of the Bogomol'nyi type 
bound in the lowest order.  In the present formulation 
the gyromagnetic ratio of the stringy 
probe particle is automatically determined to be $g=2$ which 
is the same value as the background soliton. As a result we 
can confirm the force balance via the gravitational 
spin-spin interaction. 
\end{abstract}
\vskip1cm


\section{Introduction}

Probe technique is useful for studying the approximate dynamics 
of multi-soliton systems as multi-black-holes. By the term 
of `probe' we mean a sort of collective coordinate like the centre of 
mass. So-called test particle is one of probes and is governed by 
the geodesic equation. The test particle is frequently used to
estimate the gravitational wave from binary systems in an early stage.  

Among of the multi-soliton solution, BPS multi-soliton 
is an important object.  
The existence is guaranteed by the so-called force balance. The
simplest example is the Majumdar-Papapetrou(MP) solution\cite{MP} 
in the $D=4$ Einstein-Maxwell theory or $N=2$ $D=4$
supergravity\cite{Tod}.  It is 
well known that the electrostatic force balances with the 
gravitational attractive force.  
Furthermore, the 
force balance still holds up to the gravitational 
spin-spin and dipole-spin 
interaction for the Israel-Wilson-Perjes(IWP) solution \cite{IWP} which 
becomes the MP solution in the static limit, that is, 
the IWP solution is the spinning version of the MP solution\cite{KT}. 
The force balance
condition yields the saturation of the Bogomol'nyi bound and the 
correct gyromagnetic ratio $g=2$ of a spinning charged test body\cite{KT}. 
We realise, however, that the same 
situation does not hold in the Einstein-Maxwell-Dilaton (EMD) system, that 
is, the force balance condition does not give the correct gyromagnetic
ratio\cite{Tess}. It was shown that the gyromagnetic ratio of the test body is 
different from that of the background space-time. This fact may 
indicate the non-existence of the spinning multi-soliton solution in the EMD 
system. On the other hand, we are aware of an exact solution of 
spinning 
multi-soliton in the $D=4$ Einstein-Maxwell-Dilaton-Axion (EMDA) system
\cite{susy}. Obviously the new essential ingredient is   
the axion field. Hence,  taking account of the contribution from the axion 
field is important to resolve the above discrepancy of the 
gyromagnetic  ratio. 

In this paper we derive the equation of motion(EOM) for the centre of mass of 
a string, say `stringy probe particle', coupled with the general 
metric and the 
axion field. There may be two different ways to the end at first glance.  
One of them is the Papapetrou procedure\cite{Papa} refined by 
Dixon\cite{WD} and Wald\cite{Wald}. This approach is based on the 
local conservation low for the energy-momentum tensor and current of 
gauge fields. We expect 
that one can derive the EOM for the center of the mass of a string in 
this procedure although we will not show here. 
In this paper, we will adopt another approach which is 
expected to be simpler than the Papapetrou-Dixon-Wald procedure.  

The $D=4$ EMDA system describes a low energy string theory. 
The low energy string effective action originally 
comes from the requirement of the conformal 
invariance on the two dimensional $\sigma$ 
model\cite{effective}. It tells us that 
we should back to the $\sigma$ model to derive the EOM for a stringy probe 
particle. The contribution from the electromagnetic and dilation field can be 
obtained by a certain dimensional reduction and conformal 
transformation from the string to the Einstein frame.  

The rest of this paper is organised as follows. In Sec. II, we derive 
the EOM for a stringy probe beginning with the two-dimensional 
$\sigma$ model. The curved target 
space-time is ten dimensional. Through the adequate dimensional reduction 
to four dimensions we obtain the EOM in the string frame. In this
formulation the gyromagnetic ratio of a probe is uniquely determined
to be $g=2$. In Sec. III we will give an argument of the 
energy bound, so-called 
Bogomol'nyi type bound. 
In Sec. IV, we will discuss the force balance between a spinning exact 
solution of the $D=4$ EMDA system 
and a stringy probe particle via the gravitational spin-spin 
interaction. 
We confirm that the force balance holds when the Bogomol'nyi bound 
is saturated. Finally, we will give a summary and discussion in Sec. V. 

\section{Stringy Probe Particle}

\subsection{Stringy Probe Particle in Ten Dimensions}

The action of the string coupled with the general metric and 
axion field is 
%
\begin{eqnarray}
S_p=-\frac{1}{2\pi}\int d \tau d\sigma {\sqrt {-{\rm det}(g_{ab})}}
[G_{MN}(X)g^{ab}+{\hat B}_{MN}(X)\epsilon^{ab}]\partial_a X^M 
\partial_b X^N,\label{eq:action}
\end{eqnarray}
%
where  $M=0,1,...,9$ and $a=0,1$. $g_{ab}$, $G_{MN}$ and ${\hat
B}_{MN}$ are the metric of the world volume of the string, 
the metric of the ten-dimensional space-time and 
the anti-symmetric tensor, respectively. In the 
above we omitted the dilaton term because of its small contribution. 
To obtain the effective action for the centre of the mass of a string, 
we decompose $X^M(\tau, \sigma)$ into the centre of mass and the 
excitation around it as follows,  
%
\begin{eqnarray}
X^M(\tau, \sigma)={\overline X}^M(\tau)+x^M(\tau, \sigma),\label{eq:coor} 
\end{eqnarray}
%
where ${\overline X}^M(\tau):=(1/\pi)\int_0^\pi d \sigma X^M(\tau,
\sigma) $. Inserting Eq. (\ref{eq:coor}) into Eq. (\ref{eq:action}) 
and integrating out of the coordinate $\sigma$, we obtain the 
effective action for ${\overline X}^\mu (\tau)$, 
%
\begin{eqnarray}
S_p & = & \frac{1}{2}\int d \tau
\Bigl[ G_{MN}({\overline X}){\dot {\overline X}}^M 
{\dot {\overline X}}^N +2 G_{MN} \langle \partial_a x^M 
\partial^a x^N \rangle +2\partial_I G_{MN}({\overline X})
{\dot {\overline X}}^M \langle x^I {\dot x}^N \rangle
\nonumber \\
& &  +\frac{1}{2}\partial_I \partial_J G_{MN}({\overline X})
\langle x^I x^J \rangle  {\dot {\overline X}}^M 
{\dot {\overline X}}^N +2{\hat B}_{MN}({\overline X}) \langle {\dot x}^M 
{x^N}' \rangle +2\partial_I {\hat B}_{MN}({\overline X}){\dot {\overline X}}^M
\langle x^I {x^N}' \rangle +\cdots \Bigr] \label{eq:lag}\\
& =:& \frac{1}{2}\int d\tau L_{\rm eff}({\overline X}, 
{\dot {\overline X}},t),
\end{eqnarray}
%
where
%
\begin{eqnarray}
\langle F(\tau,\sigma)\rangle := \frac{1}{\pi} \int_0^\pi 
d\sigma F(\tau, \sigma).
\end{eqnarray}
%
Here we assumed that the typical scale of the curvature of the background 
space-time is much larger than that of the typical scale of the
string. For our present purpose it is enough that the excitation can be 
approximately described in the flat metric. The expression is 
given by  
%
\begin{eqnarray}
x^M(\tau, \sigma)=\frac{i}{2}\sum_{n \neq 0}[
\alpha_n^M e^{-2in(\tau -\sigma)}+{\tilde \alpha}_n^M 
e^{-2in(\tau+\sigma)} ] \label{eq:excit}
\end{eqnarray}
%
for a closed string and is inserted into Eq. 
(\ref{eq:lag}). 

From the Euler-Lagrange equation for ${\overline X}^\mu$,
%
\begin{eqnarray}
\frac{\partial L_{\rm eff}}{\partial {\overline X}^\mu}-
\partial_\tau \Bigl( \frac{\partial L_{\rm eff}}{\partial {\dot
{\overline X}}^\mu}  \Bigr)=0,
\end{eqnarray}
%
we obtain the formal EOM for the centre of the mass, 
%
\begin{eqnarray}
& & \frac{D^2{\overline X}^M}{d\tau^2}+\Gamma^M_{IJ}n^{IJ}
+\frac{1}{2}(\partial_I \Gamma^M_{JN}
-\partial_N \Gamma^M_{IJ}){\dot {\overline X}}^J 
{\cal S}^{I N}+ \partial_I \Gamma^M_{JN}{\dot {\overline
X}}^J  N^{IN}\nonumber \\
& & ~~~~~~~~~-\frac{1}{2}{\hat H}^{M}_{~IJ}
m^{IJ}-\frac{1}{4}(\partial^M {\hat H}_{IJN}
+\partial_J {\hat H}^M_{~IN} ){\dot {\overline X}}^J 
{\cal  T}^{IN}+\cdots=0, \label{eq:first}
\end{eqnarray}
%
where ${\hat H}_{IJK}:=3\partial_{[I}{\hat B}_{JK]}$ and 
%
\begin{eqnarray}
M^{IJ}=\frac{i}{2}\sum_{n \neq 0}\frac{1}{n}(
\alpha_n^I{\tilde \alpha}_n^J - {\tilde \alpha}_n^I
\alpha_n^J)e^{-4in \tau}, ~~
N^{IJ}=\frac{i}{2}\sum_{n \neq 0}\frac{1}{n}(
\alpha_n^I{\tilde \alpha}_n^J + {\tilde \alpha}_n^I
\alpha_n^J)e^{-4in \tau} \nonumber 
\end{eqnarray}
%
%
\begin{eqnarray}
m^{IJ}=2 \sum_{n \neq 0}(
\alpha_n^I{\tilde \alpha}_n^J - {\tilde \alpha}_n^I
\alpha_n^J)e^{-4in \tau}, ~~
n^{IJ}=2 \sum_{n \neq 0}(
\alpha_n^I{\tilde \alpha}_n^J + {\tilde \alpha}_n^I
\alpha_n^J)e^{-4in \tau} \nonumber 
\end{eqnarray}
%
%
\begin{eqnarray}
{\cal S}^{IJ}=S^{IJ}+{\tilde S}^{IJ}, ~~
{\cal T}^{IJ}=-S^{IJ}+{\tilde S}^{IJ}+M^{IJ} \nonumber 
\end{eqnarray}
%
%
\begin{eqnarray}
S^{IJ}=i \sum_{n=1}\frac{1}{n}(\alpha_n^I
 \alpha_{-n}^J-\alpha_n^J \alpha_{-n}^I ), ~~
 {\tilde S}^{IJ}=i \sum_{n=1}\frac{1}{n}({\tilde \alpha}_n^I
 {\tilde \alpha}_{-n}^J-{\tilde \alpha}_n^J 
{\tilde \alpha}_{-n}^I )\nonumber 
\end{eqnarray}
%
$S^{IJ}$ and ${\tilde S}^{IJ}$ are 
the angular momentum of the right-moving and left-moving 
excitations, respectively. 
For the later comparison we assume that the left-moving excitations is 
absent, that is, 
%
\begin{eqnarray}
{\tilde \alpha}_n^M=0, \label{eq:left}
\end{eqnarray}
%
and this implies 
%
\begin{eqnarray}
n^{IJ}=m^{IJ}=M^{IJ}=N^{IJ}={\tilde S}^{IJ}=0. 
\end{eqnarray}
%
This circumstance corresponds to that of the soliton which 
will be discussed later in Sec. IV. 

As a result, Eq. (\ref{eq:first}) can be simplified in the form 
%
\begin{eqnarray}
\frac{D^2{\overline X}^M}{d\tau^2}+\frac{1}{2}(\partial_I
\Gamma^M_{JN}
-\partial_N\Gamma^M_{JI}){\dot {\overline X}}^J
S^{IN}+\frac{1}{4}(\partial^M
{\hat H}_{IJN}+\partial_J {\hat H}^M_{~NI})
{\dot {\overline X}}^J S^{IN}+\cdots =0. 
\end{eqnarray}
%
in ten dimensions. 
The second term may come from the ${}^{(10)}R^M_{~JIN}{\dot {\overline
X}}^J S^{IN}$ and stands for the gravitational 
spin-spin interaction. 
The third term is just what we wanted and expresses the axion-spin interaction
term occured due to the ``additional ingredient'', axion field. 
As we will show later, the term is essential for the force balance in the 
respect with the spin-spin interactions. We expect that the equation 
should be derived by the Papapetrou-Dixon-Wald approach too. Beside 
the axion term, the above equation is basically identical with 
the higher-dimensional 
version of one obtained in \cite{Papa}\cite{WD}\cite{Wald}.

\subsection{Stringy Probe Particle in Four Dimensions}

Next, we derive the EOM for a 
stringy probe particle in four dimensions. To this 
end we now remember that the four dimensional low energy 
effective string action for the metric, electromagnetic, axion field 
and dilaton fields can be derived by a certain dimensional reduction of 
the ten dimensional low energy string action,   
%
\begin{eqnarray}
S_{10}=\int d^{10} x e^{-2\phi}{\sqrt {-G}}
\Bigl[R_G+4G^{MN}\partial_M\phi \partial_N \phi -\frac{1}{12}
G^{MN}G^{IJ}G^{KL}{\hat H}_{MIK} {\hat H}_{NJL}\Bigr].
\end{eqnarray}
%
The actual procedure is given by\cite{dim} 
%
\begin{eqnarray}
G_{\mu\nu}=g_{S\mu\nu}+A_\mu A_\nu, ~~G_{4 \mu}=A_\mu, ~~
G_{44}=1,~~G_{ab}=\delta_{ab}, ~~G_{4a}=G_{\mu a}=0 \label{eq:trans1}
\end{eqnarray}
%
and
%
\begin{eqnarray}
{\hat B}_{\mu\nu}=B_{\mu\nu},~~{\hat B}_{4\mu}=A_\mu, \label{eq:trans2}
\end{eqnarray}
%
where $\mu=0,1,2,3$ and $a=5,6,7,8,9$. In the above we imposed that 
all of the above fields do not depend on the coordinates $x^4 \sim x^9$. 
We obtain the action in the string frame, 
%
\begin{eqnarray}
S_4=\int d^4x e^{-2\phi}{\sqrt {-g_S}}\Bigl[R_S+4g_S^{\mu\nu}\partial_\mu \phi
\partial_\nu \phi-\frac{1}{12}e^{-4\phi}g_S^{\mu\nu}g_S^{\alpha\beta}
g_S^{\rho\sigma}H_{\mu\alpha\rho}H_{\nu\beta\sigma}
-\frac{1}{2}e^{-2\phi}g_S^{\mu\nu}g_S^{\alpha\beta}
F_{\mu\alpha}F_{\nu\beta}\Bigr],
\end{eqnarray}
%
where 
$H_{\mu\nu\alpha}=3\partial_{[\mu}B_{\nu\alpha]}
+3A_{[\mu}F_{\nu\alpha]}$. To obtain the action in the Einstein frame 
we have to take the conformal transformation as
%
\begin{eqnarray}
g_{S\mu\nu}=e^{2\phi}g_{\mu\nu}. \label{eq:con}
\end{eqnarray}
%
Then we obtain the action of the EMDA system, 
%
\begin{eqnarray}
S_4=\int d^4 x {\sqrt {-g}}\Bigl[R-2g^{\mu\nu}\partial_\mu \phi 
\partial_\nu \phi-\frac{1}{12}g^{\mu\nu}g^{\alpha\beta}g^{\rho\sigma}
e^{-4\phi}H_{\mu\alpha\rho}H_{\nu\beta\sigma}
+\frac{1}{2}e^{-2\phi}g^{\mu\nu}g^{\alpha\beta}F_{\mu\alpha}
F_{\nu\beta} \Bigr].
\end{eqnarray}
%

Now we can write down the EOM in terms of the four-dimensional
quantities defined by Eqs. (\ref{eq:trans1}) and (\ref{eq:trans2}).  
In the string frame, the EOM for a stringy probe particle is 
immediately written as  
%
\begin{eqnarray}
& & \partial_\tau^2{\overline X}^\mu
+{}^{(4)}\Gamma^\mu_{\alpha\beta}({\overline X}){\dot {\overline X}}^\alpha
 {\dot {\overline X}}^\beta+({\dot {\overline X}}^4+A_\alpha {\dot
{\overline X}}^\alpha ){\dot {\overline X}}^\nu F_{\nu}^{~\mu}
-\frac{1}{2}\partial^\mu F_{\alpha\beta}({\dot {\overline
X}}^4 +A_\rho {\dot {\overline X}}^\rho )S^{\alpha\beta} \nonumber \\
& & ~~~~+\frac{1}{2}(
\partial_\alpha {}^{(4)}\Gamma^\mu_{\nu\beta}-
\partial_\beta {}^{(4)}\Gamma^\mu_{\nu\alpha} ){\dot {\overline
X}}^\nu S^{\alpha\beta}-\frac{1}{4}(\partial^\mu H_{\alpha\nu\beta}
-\partial_\nu H^\mu_{~\alpha\beta}){\dot {\overline X}}^\nu S^{\alpha\beta}
+O\Bigl( \frac{1}{r^5}\Bigr) =0. 
\end{eqnarray}
%
In the above expression, we can show that 
${\dot {\overline X}}^4+A_\alpha {\dot {\overline X}}^\alpha  $ 
is approximately conserved. Noting the equation for ${\overline X}^4$,
%
\begin{eqnarray}
\partial_\tau^2 {\overline X}^4-F_{\mu\nu} A^\nu ({\dot {\overline
X}}^4 +A_\alpha {\dot {\overline X}}^\alpha ){\dot {\overline X}}^\mu 
+{}^{(4)}\nabla_\mu A_\nu {\dot {\overline X}}^\mu {\dot {\overline
X}}^\nu=O\Bigl(\frac{1}{r^4}\Bigr), 
\end{eqnarray}
%
we can see that 
%
\begin{eqnarray}
\partial_\tau ( {\dot {\overline X}}^4+A_\mu {\dot
{\overline X}}^\mu ) = O\Bigl( \frac{1}{r^4}\Bigr)
\end{eqnarray}
%
holds and we set  
%
\begin{eqnarray}
{\dot {\overline X}}^4+A_\mu {\dot {\overline X}}^\mu
=\frac{q}{m}+O\Bigl(\frac{1}{r^4}\Bigr).
\end{eqnarray}
%
Finally we are resulted in 
%
\begin{eqnarray}
& & \partial_\tau^2{\overline X}^\mu
+{}^{(4)}\Gamma^\mu_{\alpha\beta}({\overline X}){\dot {\overline X}}^\alpha
 {\dot {\overline X}}^\beta+\frac{q}{m}{\dot {\overline X}}^\nu F_{\nu}^{~\mu}
-\frac{q}{2m}\partial^\mu F_{\alpha\beta}S^{\alpha\beta} \nonumber \\
& & ~~~~+\frac{1}{2}(
\partial_\alpha {}^{(4)}\Gamma^\mu_{\nu\beta}-
\partial_\beta {}^{(4)}\Gamma^\mu_{\nu\alpha} ){\dot {\overline
X}}^\nu S^{\alpha\beta}-\frac{1}{4}(\partial^\mu H_{\alpha\nu\beta}
-\partial_\nu H^\mu_{~\alpha\beta}){\dot {\overline X}}^\nu S^{\alpha\beta}
+O\Bigl( \frac{1}{r^5} \Bigr) =0 \label{eq:master}
\end{eqnarray}
%
in four dimensional string frame. 
Note that we can read from the above expression 
that the gyromagnetic ratio of the string probe particle is $g=2$. In
the covariant form we may expect that the equation 
%
\begin{eqnarray}
\frac{{\cal D}^2{\overline X}^\mu}{d\tau^2}-\frac{q}{m}F^\mu_{~\nu}
{\dot {\overline X}}^\nu
-\frac{q}{2m}{}^{(4)}\nabla^\mu F_{\alpha\beta}S^{\alpha\beta}
+\frac{1}{2}{}^{(4)}R^\mu_{~\nu\alpha\beta}{\dot {\overline X}}^\nu
S^{\alpha\beta}-\frac{1}{4}\Bigl( {}^{(4)}\nabla^\mu
H_{\alpha\nu\beta} -{}^{(4)}\nabla_\nu H^\mu_{~\alpha\beta}  \Bigr)
{\dot {\overline X}}^\nu S^{\alpha\beta} +\cdots =0
\end{eqnarray}
%
holds, where ${\cal D}/d\tau := {\dot {\overline X}}^\mu
{}^{(4)}\nabla_\mu$. 

We can easily see that the dilaton field has no contributions 
to the leading order of the spin-spin interaction\cite{Tess}. Since we are 
only interested in the force balance via the spin-spin interaction in 
this paper, we can go on in the string frame. If one wants to 
study the force balance in the Einstein frame, 
one should insert the above conformal
transformation(Eq.(\ref{eq:con})) into Eq. (\ref{eq:master}). 
It is, however, well known that the force balance between 
monopole components yields the saturation of a Bogomol'nyi bound
($|Q|={\sqrt {2}}M$, in the present stringy case).

\section{Energy Bound for D=4 Einstein-Maxwell-Dilaton-Axion Theory}

In this section, we give the energy bound argument 
for the $D=4$ EMDA system in 
the Einstein frame. 
This is basically same as the proof given in Ref. \cite{BB} 
except for the existence of the axion field. 
In four dimensions the axion field can be expressed by a scalar field, 
$a$, as follows,  
%
\begin{eqnarray}
H_{\mu\nu\alpha}=-e^{4\phi}\epsilon_{\mu\nu\alpha\beta}\partial^\beta
a.
\end{eqnarray}
%
%
%
Here, bearing $N=4$ $D=4$ supergravity in mind, 
we define the supercovariant derivative for a spinor $\epsilon$ by  
%
\begin{eqnarray}
{\hat \nabla}_\mu \epsilon=\nabla_\mu \epsilon
-\frac{i}{4}e^{2\phi}\partial_\mu a \epsilon+\frac{i}{4{\sqrt
{2}}}e^{-\phi}\gamma^{\rho\sigma}\gamma_\mu F_{\rho\sigma}\epsilon, 
\end{eqnarray}
%
where 
$\nabla_\mu\epsilon:=(\partial_\mu+\Gamma_\mu)\epsilon$ and $\Gamma_\mu$ 
is the spin connection, 
%
\begin{eqnarray}
\Gamma_\mu=-\frac{1}{8}e^{\nu {\hat k}} \nabla_\mu e_\nu^{{\hat \ell}}
[\gamma_{{\hat \ell}}, \gamma_{{\hat k}}]. 
\end{eqnarray}
%
In this section $\nabla_\mu$ stands for the covariant derivative associated
with the Einstein metric $g_{\mu\nu}$. 
The spinor $\epsilon$ is assumed to satisfy the modified Witten
equation $\gamma^i {\hat \nabla}_i \epsilon=0$.  
Furthermore, we define the Nester-like tensor\cite{Nester}, 
%
\begin{eqnarray}
{\hat E}^{\mu\nu}& := & \frac{1}{2}\Bigl({\overline 
   \epsilon}\gamma^{\mu\nu\alpha}{\hat \nabla}_\alpha \epsilon
-{\overline {{\hat \nabla}_\alpha \epsilon}}\gamma^{\mu \nu \alpha}
   \epsilon\Bigr) \nonumber \\
& = & E^{\mu\nu}-\frac{i}{4}e^{2\phi}\partial_\alpha a 
{\overline \epsilon}\gamma^{\mu\nu\alpha} \epsilon
-\frac{i}{{\sqrt {2}}}e^{-\phi}{\overline
   \epsilon}(F^{\mu\nu}-\gamma_5
{\tilde F}^{\mu\nu})\epsilon, 
\end{eqnarray}
%
where ${\tilde
F}^{\mu\nu}=(1/2)\epsilon^{\mu\nu\alpha\beta}F_{\alpha\beta}$. 
Hereafter we assume that $\epsilon$ has the chirality 
of $\gamma_5 \epsilon = i \epsilon$. This is influenced by $N=4$ 
$D=4$ supergravity. 
The divergence of the Nester-like tensor becomes
%
\begin{eqnarray}
\nabla_\nu {\hat E}^{\mu\nu}& 
= & \frac{1}{2}G^{\mu}_\nu V^\nu+{\overline {{\hat \nabla}_\alpha
\epsilon}}\gamma^{\mu\nu\alpha}{\hat \nabla}_\beta \epsilon
-\frac{1}{2}T^\mu_\nu(F)V^\nu-\frac{1}{{\sqrt {2}}}e^{\phi}\nabla_\nu
(e^{-2\phi}F^{\mu\nu}){\overline \epsilon}\epsilon \nonumber \\
& & - \frac{i}{2}\partial_\nu \phi\partial_\alpha a e^{2\phi}{\overline \epsilon}
\gamma^{\mu\nu\alpha}\epsilon-\frac{i}{{\sqrt {2}}}
e^\phi \partial_\nu \phi {\overline \epsilon}(F^{\mu \nu}
+\gamma_5{\tilde F}^{\mu\nu})\epsilon,
\end{eqnarray}
%
where
%
\begin{eqnarray}
T_{\mu\nu}(F)=e^{-2\phi}\Bigl(
F_{\mu\rho}F_\nu^\rho-\frac{1}{4}g_{\mu\nu} F^2\Bigr).
\end{eqnarray}
%
After some tedious calculations we obtain
%
\begin{eqnarray}
\nabla_\nu {\hat E}^{\mu\nu}=T^\mu_\nu({\rm mat})V^\nu +
{\overline {{\hat \nabla}_\alpha
\epsilon}}\gamma^{\mu\alpha\beta}{\hat \nabla}_\beta \epsilon
-{\overline {\delta \lambda}} \gamma^\mu \delta \lambda 
+\frac{i}{{\sqrt {2}}}{\overline \epsilon}J^\mu ({\rm mat}) \epsilon 
\label{eq:basic},
\end{eqnarray}
%
where $V^\mu={\overline \epsilon} \gamma^\mu \epsilon$ and 
$\delta \lambda$ is defined by
%
\begin{eqnarray}
\delta \lambda =\frac{1}{{\sqrt {2}}}\Bigl( \gamma^\mu \partial_\mu 
\phi+\frac{1}{2}e^{2\phi}\gamma^\mu \gamma_5 \partial_\mu a 
-\frac{i}{2{\sqrt {2}}} \gamma^{\mu \nu} F_{\mu \nu}\Bigr)\epsilon.
\end{eqnarray}
%
One can interpret the field $\lambda$ as the dilatino
which is a superpartner of the dilation field in supergravity. 
In the derivation of Eq. (\ref{eq:basic}) we used the field equations 
%
\begin{eqnarray}
G_{\mu\nu}=2T_{\mu\nu}(\phi)+2T_{\mu\nu}(a)+2T_{\mu\nu}(F)
+2T_{\mu\nu}({\rm mat})
\end{eqnarray}
%
and
%
\begin{eqnarray}
\nabla_\nu (e^{-2\phi}F^{\mu\nu})+\frac{1}{2}{\tilde F}^{\mu\nu}
\partial_\nu a =-J^\mu({\rm mat}),
\end{eqnarray}
%
where 
%
\begin{eqnarray}
T_{\mu\nu}(\phi)=\partial_\mu\phi\partial_\nu \phi-\frac{1}{2}
g_{\mu\nu}\partial^\rho \phi \partial_\rho \phi
\end{eqnarray}
%
and
%
\begin{eqnarray}
T_{\mu\nu}(a)=\frac{e^{4\phi}}{4}\Bigl( 
\partial_\mu a\partial_\nu a-\frac{1}{2}
g_{\mu\nu}\partial^\rho a \partial_\rho a \Bigr).
\end{eqnarray}
%
We also used the following expression, 
%
\begin{eqnarray}
{\overline {\delta \lambda}} \gamma^\mu \delta \lambda 
& = & T^\mu_\nu (\phi) V^\nu -T^\mu_\nu
(a)V^\nu-\frac{1}{2}T^\mu_\nu(F)
-\frac{1}{2}e^{2\phi}\partial_\nu \phi \partial_\alpha a {\overline \epsilon}
\gamma^{\nu \mu \alpha}\gamma_5 \epsilon \nonumber \\
& & +\frac{i}{{\sqrt
{2}}}e^{-\phi}
\partial_\nu \phi {\overline \epsilon}(F^{\mu\nu}+\gamma_5
{\tilde F}^{\mu\nu})\epsilon +\frac{i}{2{\sqrt
{2}}}e^{-\phi} \partial_\nu a {\overline \epsilon}(F^{\mu\nu}+\gamma_5
{\tilde F}^{\mu\nu})\gamma_5 \epsilon.  
\end{eqnarray}
%
Integrating Eq. (\ref{eq:basic}) over the full space volume $V$ and 
assuming the dominant energy condition, we obtain the inequality   
%
\begin{eqnarray}
\int_{S_\infty} dS_{\mu\nu}{\hat E}^{\mu\nu}& = & \int_V d \Sigma_\mu 
\nabla_\nu {\hat E}^{\mu\nu} \nonumber \\ 
& = & \int_V d\Sigma_\mu 
\Bigl[ T^\mu_\nu({\rm mat})V^\nu+
{\overline {{\hat \nabla}_\alpha
\epsilon}}\gamma^{\mu\alpha\beta}{\hat \nabla}_\beta \epsilon
-{\overline {\delta \lambda}} \gamma^\mu \delta \lambda 
+\frac{i}{{\sqrt {2}}}{\overline \epsilon}J^\mu ({\rm mat}) \epsilon
\Bigr] \geq 0. 
\end{eqnarray}
%
The left hand-side in the first line of the above can be written in the
term of several conserved charges, ADM four-momentum $P^\mu_{\rm ADM}$, 
the electric charge $Q$ and magnetic charge $P$, as follows\cite{axion},  
%
\begin{eqnarray}
\int_{S_\infty} dS_{\mu\nu}{\hat E}^{\mu\nu}= {\overline \epsilon}_0 
\gamma_\mu \epsilon_0 P^\mu_{\rm ADM}-\frac{1}{{\sqrt {2}}}
 {\overline \epsilon}_0 (Q-\gamma_5 P) \epsilon_0,  
\end{eqnarray}
%
where $\epsilon_0$ is the constant spinor which is identical to the 
limit value of $\epsilon$ at the spatial infinity. 
Hence we obtain the Bogomol'nyi type bound, 
%
\begin{eqnarray}
M \geq \frac{1}{{\sqrt {2}}}{\sqrt {Q^2+P^2}}.
\end{eqnarray}
%
In the absence of the magnetic charge, $P=0$, the saturation of the 
bound occurs as $M=(1/{\sqrt {2}})|Q|$. 

\section{Force Balance via Gravitational Spin-Spin Interaction}

In this section we will confirm the force balance between the 
gravitational spin-spin, 
dipole-spin and axion-spin interactions when the Bogomol'nyi type bound is 
saturated. As we said in Sec. I, the spinning multi-soliton   
solution exists in the $D=4$ EMDA system. This solution is the 
straightforward extension of the IWP solution and saturates the
Bogomol'nyi bound, {\it i.e.}, $|Q|={\sqrt {2}}M$. We expect that the force
balance holds via the gravitational spin-spin interaction therein. To 
check the force balance, we consider the motion of 
a stringy probe particle on the background geometry of the single 
soliton with $Q={\sqrt {2}}M$. In the Einstein frame 
the single solution is given by\cite{susy} 
%
\begin{eqnarray}
ds^2=-e^{2\phi}(dt+\omega_idx^i)^2+e^{-2\phi}d{\bf x}^2\label{eq:metric1}
\end{eqnarray}
%
and
%
\begin{eqnarray}
A=\frac{1}{{\sqrt {2}}}e^{2\phi}(dt+\omega_idx^i), \label{eq:metric2}
\end{eqnarray}
%
where $i=1,2,3$ and $\epsilon^{ijk}\partial_j \omega_k=-\partial_i a$. 
In the limit of the slow rotating the metric, vector potential and 
axion field are written as 
%
\begin{eqnarray}
ds^2 =-\Bigl(1+\frac{2M}{r}\Bigr)^{-1}dt^2+\Bigl(1+\frac{2M}{r}\Bigr)
d{\bf x}^2+\frac{4\epsilon_{ijk}J^jx^k}{r^3}dtdx^i +O(\alpha^2),
\end{eqnarray}
%
%
\begin{eqnarray}
A\simeq\frac{1}{{\sqrt {2}}}e^{2\phi}dt-\frac{Q}{r^3M}\epsilon_{ijk}
J^jx^kdx^i 
\end{eqnarray}
%
and
%
\begin{eqnarray}
a=\frac{2J^i x^i}{r^3}+O(\alpha^2), 
\end{eqnarray}
%
where $\alpha$ is the parameter of the angular momentum so that 
$\vec{J}=(0,0,M\alpha)$. One can find the exact expression of the 
above solution in Ref. \cite{susy}. 

In the order of $O(1/r^3)$, the spatial components of the  
equation of the motion for a probe is given by 
%
\begin{eqnarray}
\partial_\tau^2 {\overline X}^i & \simeq & -{}^{(4)}\Gamma^i_{\alpha\beta}
{\dot {\overline X}}^\alpha {\dot {\overline X}}^\beta 
+\frac{q}{m}F^i_{~\alpha}{\dot {\overline X}}^\alpha \nonumber \\
& \simeq &  -{}^{(4)}\Gamma^i_{00}+\frac{q}{m}F^i_{~0} \nonumber \\
& = & \frac{1}{2}g_S^{ij}\partial_j g_{S 00}+\frac{q}{m}\partial^i A_0
\nonumber \\
&\simeq & 
-\Bigl( 2-{\sqrt {2}}\frac{q}{m} \Bigr)\partial^i \phi   
\end{eqnarray}
%
because of $g_{S00}=e^{2\phi}g_{00}=-e^{4\phi}$. 
Hence we can see the force cancellation between 
the Newtonian-like gravitational and the Coulombic electrostatic
forces when $q={\sqrt {2}}m$ holds. 

Let us confirm the force balance via the gravitational 
spin-spin interaction.  
The spin-spin, dipole-spin and axion-spin interaction forces are 
%
\begin{eqnarray}
F^i_{\rm spin}=-\frac{1}{2}R^i_{~0jk}S^{jk}+O\Bigl(\frac{1}{r^5} \Bigr)
\end{eqnarray}
%
%
\begin{eqnarray}
F^i_{\rm dipole}=\frac{q}{2m}S^{jk}{}^{(4)}\nabla^i F_{jk}
+O\Bigl(\frac{1}{r^5} \Bigr)
\end{eqnarray}
%
and
%
\begin{eqnarray}
F^i_{\rm axion}=-\frac{1}{4}\partial^i H_{j0k}S^{jk}
+O\Bigl(\frac{1}{r^5} \Bigr),
\end{eqnarray}
%
respectively. Thus the total force in the order of $O(1/r^4)$ becomes 
%
\begin{eqnarray}
F^i_{\rm spin}+F^i_{\rm dipole}+F^i_{\rm axion}
=\Bigl(2-\frac{qQ}{mM} \Bigr)\partial_i 
\Bigl[ \frac{-{\bf J}\cdot{\bf S}+3({\bf S}\cdot{\hat {\bf r}})
({\bf J}\cdot {\hat {\bf r}})  }{r^3}   \Bigr]+O\Bigl(\frac{1}{r^5} \Bigr),
\end{eqnarray}
%
where $S^i:=(1/2)\epsilon^{ijk}S^{jk}$ and ${\hat {\bf r}}:={\bf r}/r$. 
As we can easily see, the above total force exactly vanishes when the 
Bogomol'nyi type bound is saturated, that is, $Q={\sqrt {2}}M$ and 
$q={\sqrt {2}}m$. 

\section{Summary and Discussion}

In this paper we derived the equation of motion for a stringy probe 
particle coupled with the axion field. Its gyromagnetic ratio is 
automatically determined in the present formulation and turns out to 
be $g=2$. Moreover, we confirmed that 
the force balance holds via the gravitational 
spin-spin interaction when the Bogomol'nyi type bound is saturated. The 
behaviour of the probe seems to be related to the existence of the 
multi-soliton solution in the $D=4$ 
Einstein-Maxwell-Dilaton-Axion(EMDA) system.

We investigated the force balance in four dimensions. 
As we can see the multi-soliton solution has naked singularity 
in the $D=4$ EMDA system\cite{susy} as well as the 
Kerr-Newman space-time at the saturation of the Bogomol'nyi bound. 
On the other hand, 
Horowitz and Sen found a multi-soliton solution without naked singularity 
in $D>5$-dimension\cite{HS}. They also made 
the four dimensional black hole space-time by taking doubly 
periodic allays in six dimensions. 
So the force balance therein is also 
important and it seems to be 
easy to extend our present work to the higher-dimensional version. 

As we moderately said, the value of the gyromagnetic ratio is 
important to discover exact spinning multi-soliton solutions. Before 
we try to discover it is advisable to check the value of probe gyromagnetic
ratio, that is, whether the value gyromagnetic ratio is 
same as that of the background soliton or not. If same, we are able to 
have a great chance to discover a new solution. 
We think that this phrase may be valid 
for higher-dimensional objects like D-brane\cite{Pol}
\cite{brane}. In addition, 
one might be going to find an exact solution of the cosmological 
spinning multi-black hole which should be the extension of the 
cosmological non-spinning multi-black hole exact solution discovered by 
Kastor and Traschen\cite{KT2}. So, we should check the force 
balance in cosmos. 

Our basic equation (Eq. (\ref{eq:master})) might be useful to study the 
approximate dynamics of a test string in a curved background 
space-time. Taking account of the axion terms is essential if the 
string is excited. If the string stays in the ground state, the 
motion is simply geodesic one. In our approach, a test string is
assumed to be widely separated from the central soliton. If not so, 
we cannot use the excitation form of Eq. (\ref{eq:excit}) in the flat
metric and must consider the time evolution due to the strong 
gravity or tidal forces\cite{teststring}.  

Finally, we should give a comment on the assumption of Eq. 
(\ref{eq:left}). From the analysis given in Sec. IV, we realise that the 
stringy probe particle plays as the test body like the central 
soliton under the assumption. Namely, the soliton given by 
Eqs. (\ref{eq:metric1}) and (\ref{eq:metric2}) seems to be related to 
the right-moving string state. This is because the solution 
is in the BPS state 
and the absence of the left(or right) moving excitation(oscillation) 
is essential for the BPS state\cite{BPS}. 

Beside our present purpose of the force balance, Eq. (\ref{eq:first}) 
should be used for general cases having both excitations. The
comparison with the equation derived by the Papapetrou
procedure\cite{Papa} 
is important because it is not likely that 
the extra terms does not appear in the procedure. 

\vskip1cm

\centerline{\bf Acknowledgment}
The author is grateful to Gary Gibbons and DAMTP relativity group for their 
hospitality. He also thanks M. Spicci for his careful reading of this 
manuscript. This study is supported by JSPS(No. 310). 


\begin{thebibliography}{22}
\bibitem{MP}
S. D. Majumdar, Phys. Rev. {\bf 72}, 930(1947)\\
A. Papapetrou, Proc. Roy. Irish Acad.{\bf A51}, 191(1947)
\bibitem{Tod}
K. P. Tod, Phys. Lett. {\bf B121}, 241(1981)
\bibitem{IWP}
Z. Perjes, Phys. Rev.Lett.{\bf 27}, 1668(1971)\\
W. Israel and G. A. Wilson, J. Math. Phys.{\bf 13}, 865(1972)
\bibitem{KT}
J. Tiomno, Phys. Rev. {\bf D7}, 356(1973);\\
W. Israel and J. T. J. Spanos, Lett. Nouvo Cim. {\bf 7}, 245(1973);\\
D. Kastor and J. Traschen, Class. Quantum Grav. {\bf 16}, 1265(1999)
\bibitem{Tess}
T. Shiromizu, to be published in Phys. Lett. B(1999), hep-th/9906177
\bibitem{susy}
R. Kallosh, D. Kastor, T. Ortin and T. Torma, Phys. Rev. {\bf D50}, 
6374(1994);\\
E. Bergshoeff, R. Kallosh and T. Ortin, Nucl. Phys. {\bf 478}, 156(1996)
\bibitem{Papa}
A. Papapetrou, Proc. Roy. Soc. {\bf A209}, 248(1951)
\bibitem{WD}
W. G. Dixon, Phil. Trans. Roy. Soc. {\bf 277A}, 59(1960);\\
Proc. Roy. Soc. Lond. {\bf A314}, 499(1970)
\bibitem{Wald}
R. M. Wald, Phys. Rev. {\bf D6}, 406(1972) 
\bibitem{effective}
C. G. Callan, D. Friedan, E. J. Martinec and M. J. Perry, 
Nucl. Phys. {\bf B262}, 593(1985)
\bibitem{dim}
E. Bergshoeff, R. Kallosh and T. Ortin, Phys. Rev. {\bf D50}, 5188(1994)
\bibitem{axion}
For the typical example in the Sec. IV, the axionic charge does not 
affect the bound. In fact, the axionic charge term becomes $\sim \int
dS_{\mu\nu}\epsilon^{\mu\nu\alpha\beta}
\partial_\alpha a {\overline \epsilon}_0 \gamma_\beta \epsilon_0 
\sim \int dS (x^i/r) \epsilon^{0ijk} \partial_j a 
{\overline \epsilon}_0 \gamma_k \epsilon_0 =0 $.
\bibitem{BB}
G. W. Gibbons, D. Kastor, L. A. J. London, P. K. Townsend and 
J. Traschen, Nucl. Phys. {\bf B416}, 850(1994)
\bibitem{Nester}
J. Nester, Phys. Lett. {\bf A83}, 241(1981)
\bibitem{HS}
G. T. Horowitz and A. Sen, Phys. Rev. {\bf D53}, 808(1996)
\bibitem{Pol}
J. Polchinski, Phys. Rev. Lett., {\bf 75}, 4724(1995)
\bibitem{brane}
A. A. Tseytlin, Nucl. Phys. {\bf B487}, 141(1997);\\
M. Cevetic and D. Youm, Nucl. Phys. {\bf B499}, 253(1997);\\
V. Balasubramanian, D. Kastor, J. Traschen and K. Z. Win, 
Phys. Rev. {\bf D59}, 084007(1999)
\bibitem{KT2}
D. Kastor and J. Traschen, Phys. Rev. {\bf D47}, 5370(1993)
\bibitem{teststring}
G. T. Horowitz and A. R. Steif, Phys. Rev. {\bf D42}, 1950(1990);\\
G. T. Horowitz and S. F. Ross, Phys. Rev. {\bf D57}, 1098(1998)
\bibitem{BPS}
For example, J. M. Maldacena, PhD Thesis, hep-th/9607235
\end{thebibliography}
\end{document}